\documentclass[conference, compsocconf]{IEEEtran}
\usepackage[utf8]{inputenc}
\usepackage[english]{babel}
\PassOptionsToPackage{bookmarks=false}{hyperref}
\usepackage{multirow, multicol, booktabs, tabulary, tabu, longtable, array, varwidth}
\usepackage{placeins, lipsum}
\usepackage[flushleft]{threeparttable}

\usepackage[hyphens]{url}
\usepackage[labelfont=bf]{caption}
\usepackage{cite}
\usepackage{hyperref}
\usepackage[dvipsnames, table]{xcolor}
\hypersetup{
    linkcolor  = violet!85!black,
    citecolor  = magenta!85!black,
    urlcolor   = blue!85!black,
    colorlinks = true,
    breaklinks = true
}

\usepackage{enumerate}

\usepackage[inline]{enumitem}

\usepackage{amsmath, amsfonts, amssymb, amsthm, nicefrac}
\usepackage{siunitx}
\theoremstyle{definition}

\usepackage{listings}

\usepackage[scaled=0.82]{beramono}

\usepackage[]{cleveref} %
\crefname{appsec}{Appendix}{Appendices}
\crefformat{section}{\S#2#1#3}
\crefformat{subsection}{\S#2#1#3}
\crefformat{subsubsection}{\S#2#1#3}
\crefformat{equation}{(#2#1#3)}
\crefrangeformat{equation}{(#3#1#4--#5#2#6)}
\crefmultiformat{equation}{(#2#1#3)}{ and~(#2#1#3)}{, (#2#1#3)}{ and~(#2#1#3)}
\crefformat{figure}{Fig.~#2#1#3}
\crefrangeformat{figure}{Figs. #3#1#4--#5#2#6}
\crefmultiformat{figure}{Figs.~#2#1#3}{ and~#2#1#3}{, #2#1#3}{ and~#2#1#3}
\crefname{algocf}{Alg.}{Algs.}
\Crefname{algocf}{Algorithm}{Algorithms}
\crefformat{table}{Table~#2#1#3}
\crefrangeformat{table}{Tables~#3#1#4--#5#2#6}
\crefmultiformat{table}{Tables~#2#1#3}{ and~#2#1#3}{, #2#1#3}{ and~#2#1#3}

\usepackage{textcomp}
\usepackage[shortcuts,acronym]{glossaries}
\usepackage{soul, footnote, xargs}
\usepackage[colorinlistoftodos,prependcaption,textsize=tiny]{todonotes}
\usepackage{balance}

\usepackage{graphicx}
\usepackage{subcaption}
\usepackage{tikz, epstopdf, stfloats, bbding, capt-of}
\usepackage{algorithmic}
\usepackage[ruled, linesnumbered]{algorithm2e}

\SetCommentSty{mycommfont}
\usepackage{soul}
\sethlcolor{black}
\makeatletter
\newif\if@blind
\@blindtrue %
\if@blind \sethlcolor{black}\else
   
\fi

\begin{document}

\title{A Study of Network Congestion in Two Supercomputing High-Speed Interconnects \\
}

\author{\IEEEauthorblockN{
    {Saurabh Jha}\IEEEauthorrefmark{1},
    {Archit Patke}\IEEEauthorrefmark{1},
    {Jim Brandt}\IEEEauthorrefmark{2},
    {Ann Gentile}\IEEEauthorrefmark{2},
    {Mike Showerman}\IEEEauthorrefmark{4},
    {Eric Roman}\IEEEauthorrefmark{3},\\
    {Zbigniew T. Kalbarczyk}\IEEEauthorrefmark{1},
    {Bill Kramer}\IEEEauthorrefmark{1,4},
    and
    {Ravishankar K. Iyer}\IEEEauthorrefmark{1}
    }
    \IEEEauthorblockA{\IEEEauthorrefmark{1}{University of Illinois at Urbana-Champaign}}
    \IEEEauthorblockA{\IEEEauthorrefmark{2}{Sandia National Labs}}
    \IEEEauthorblockA{\IEEEauthorrefmark{4}{National Center for Supercomputing Applications}}
    \IEEEauthorblockA{\IEEEauthorrefmark{3}{National Energy Research Scientific Computing Center}}
}
\maketitle

\begin{abstract}
Network congestion in high-speed interconnects is a major source of application runtime performance variation. 
Recent years have witnessed a surge of interest from both academia and industry in the development of novel approaches for congestion control at the network level and in application placement, mapping, and scheduling at the system-level. 
However,  these studies are based on proxy applications and benchmarks that are not representative of field-congestion characteristics of high-speed interconnects.
To address this gap, we present (a) an end-to-end framework for monitoring and analysis to support long-term field-congestion characterization studies, and (b) an empirical study of network congestion in petascale systems across two different interconnect technologies: (i) Cray Gemini, which uses a 3-D torus topology, and (ii) Cray Aries, which uses the DragonFly topology.  
\end{abstract}

\section{Introduction}
\label{s:introduction}
Despite years of innovation in network routing, congestion avoidance, and mitigation algorithms across generation high-performance interconnects, extreme-scale applications running on high-performance computing systems continue to suffer from performance variation and scaling challenges~\cite{bhatele2013there, jain2016evaluating, hoefler2010characterizing} due to (i) frequent exposure to congestion; and (ii) the inability to automatically optimize resource parameters (such as placement, rank mapping, and application scheduling) to improve network utilization. The current interconnects suffer from congestion that can occur because of (i) bad application placement (e.g., tightly packed ranks versus ranks spread across the network)~\cite{agarwal2006topology,mubarak2017quantifying} and presence of bully applications~\cite{yang2016watch}; (ii) the presence of large numbers of failed links~\cite{jha2017resiliency} and new link failures, which force adversarial traffic shaping/flow on the network; and (iii) inherent congestion susceptibility due to network design choices (e.g., directional order routing in torus networks) or design bugs/gaps (e.g., incorrect/bad dynamic routing policy).

Growing body of work on network congestion~\cite{agarwal2006topology,mubarak2017quantifying,yang2016watch,jha2017resiliency,bhatele2013there,hoefler2010characterizing,jha2018characterizing} reflect the challenges in achieving high application performance, and the difficulty of providing timely monitoring and diagnosis of network congestion. However, the prior studies and tools are limited to proxy applications, simulations and benchmarks and hence are not reflective of production characteristics and issues. This is mainly due to inability to monitor, collect, analyze data on network congestion. To fill this gap, we provide (a) an end-to-end monitoring and analysis tool for understanding congestion causes and support long-term field-congestion characteristics, and (b) an empirical study of network congestion in high-performance computing systems. In particular, we present empirical measurements and insights obtained from two generations of Cray interconnects by using the Lightweight Distributed Metric Service (LDMS)~\cite{LDMSSC14} as a monitoring tool, and Monet~\cite{MONET:NSDI:2020} as a diagnosis tool. The studied high-speed interconnects are: (a) a Cray Gemini network deployed on Blue Waters~\cite{bluewatersweb} which uses a 3D torus-based topology and direction order routing; and (b) a Cray Aries network deployed on Edison~\cite{edison} and 2-cabinet experimental system that uses a DragonFly-based topology and dynamic routing. The study provides empirical data and insights to influence research directions in application run-time, networking and system development. 

\textit{Our contributions include the following}:
\begin{itemize}
    \item Measurements obtained from a production system running production workloads,
    \item Demonstration of end-to-end monitoring and analysis framework at scale. The proposed end-to-end framework uses LDMS~\cite{LDMSSC14} for monitoring and Monet~\cite{MONET:NSDI:2020} for diagnosis of network congestion to generate actionable insights for application developers, system managers, and network designers. LDMS collects performance-related information on links via Cray's \texttt{gpcdr}~\cite{craygpcdr} kernel module, whereas Monet uses a combination of  data science and machine learning techniques on LDMS-collected data to enable online diagnosis, data summarization, and visualization. 
    \item An empirical study of network congestion in petascale systems across two different interconnect technologies: (i) Cray Gemini, which uses a 3-D Torus topology; and (ii) Cray Aries which uses the DragonFly topology.
\end{itemize}

\textit{The key results obtained from empirical measurements are}:
\begin{itemize}
    \item Despite the use of low-level flow control and routing algorithms, hotspots in the network are common and exist for long duration of time.
    \item Heterogeneity in link bandwidth across different link-types (electrical and optical links) increases the susceptibility to congestion.
    \item Use of adaptive routing and a low-diameter network topology (such as the Cray Aries DragonFly topology) significantly improves congestion avoidance and mitigation compared to non-adaptive routing protocols, such as directional order routing in a high-diameter network topology (such as Cray Gemini Torus networks). 
    \item Identification of design gaps in HPC interconnects. Using our monitoring and analysis, we found that routing algorithms in the production network may not choose the least congested path among symmetrical paths.
\end{itemize}
More detailed results on Blue Waters Cray Gemini network can be found in~\cite{MONET:NSDI:2020}.

\section{Network and Data Description}
\label{s:data}
In this paper, we demonstrate our monitoring tool on two production systems, Blue Waters and Edison, which use the Cray Gemini and Cray Aries interconnects respectively. Subsection~\ref{subsec:sys} provides system description, Subsection~\ref{subsec:data} provides dataset description and Subsection~\ref{subsec:metric} defines the congestion metric used in the study.

\subsection{System and Interconnect Description}\label{subsec:sys}
Table~\ref{tab:system} shows differences and similarities between the two production system with respect to network interconnect technology features. 
\begin{table}[]
	\centering
	\small
	\begin{threeparttable}
		\begin{tabular}{lll}
				\toprule
				{\bf} & {\bf Blue Waters } & {\bf Edison}\\
				\midrule
				Flow control &  Credit-based & Credit-based \\
				Technology & Cray Gemini & Cray Aries \\
				Topology & 3D Torus & Dragonfly \\ 
                Routing & Directional-order routing & Adaptive \\
                Number of Nodes & 27,648 & 5,586 \\
                \bottomrule
	    \end{tabular}
	\end{threeparttable}
	\caption{\label{tab:system} Differences between production systems}
\end{table}

NCSA's (National Center for Supercomputing Applications) Blue Waters system is composed of 27,648 nodes and has a large-scale (13,824 x 48 port switches) Gemini 3D torus (dimension 24x24x24) interconnect. The available bandwidth on a particular network link is dependent on the link type (i.e., electrical vs optical) and 
number of tiles in the link. Where multiple links from a Gemini switch connect in the same direction,
it is convenient to consider them as a \emph{directionally aggreggated link} which we will
henceforth call \emph{link}, one in each of the 6 directions, X+/-, Y+/-,Z+/-, in the torus.
For large XE/XK systems~\cite{PedrettiGPCD} all such aggregated X links have an aggregate
bandwidth of 9.4 GB/s, Y links alternate between 9.4 GB/s and and 4.7 GB/s,
and Z links are predominantly 15 GB/s with 1/8 of them at 9.4 GB/s. Gemini interconnect uses directional-order routing which is predominantly static.

NERSC's (National Energy Research Scientific Computing Center) Edison system is composed of 5,586 nodes and has an (1,440 x 48 port switches) Aries DragonFly interconnect with 15 electrical groups. Electrical links (\emph{Green} and \emph{Black}) connect Aries switches within the group, where as, optical links (\emph{Blue}) form group to group connections. Optical and electrical links have a bandwidth of 1.56 GB/s and 1.75 GB/s respectively. Adaptive routing is used to route packets on non-minimal paths to alleviate congestion on the minimal path.

\subsection{Field-Congestion Datasets}\label{subsec:data}
Network performance on links is exposed via Cray's \emph{gpcdr} kernel module. These are not collected nor made available
for analysis via vendor-provided collection mechanisms. This data is collected and transported off the system for storage and analysis via the Lightweight Distributed
Metric Service (LDMS) monitoring framework~\cite{LDMSSC14}. LDMS daemons synchronize their sampling (node to node time skew not accounted for) in order to provide coherent snapshots of
network state across the whole system. The resolution of sampling is one second on Edison and sixty seconds on Blue Waters. In this work, we demonstrate the capability using one week of production data; that amounts to 7.7 TB for Edison and 370 GB for Blue Waters. 

\subsection{Congestion Metric}\label{subsec:metric}
In this paper, we use \emph{Percent Time Stalled} (\emph{PTS}) as a congestion metric to quantify network congestion. It is a suitable metric for interconnect networks that use credit-based flow control algorithms. In credit-based flow control networks, a source is allowed to send packets to the destination only if the source has sufficient credits. If sufficient credits are not available, then the link stalls, and the percentage of time spent in stalled state per unit time quantifies the extent of congestion. In this paper, we refer to this metric (i.e., percentage of time spent in stalled state per unit time) as the \emph{Percent Time Stalled}.
\section{Tool Demonstration}
\label{s:Hotspot Links}
\begin{figure*}[ht]
	\centering
	\hspace{0.03\textwidth}
	\begin{subfigure}[c]{0.46\textwidth}
    \label{fig:gemini_hotpsot}
	\includegraphics{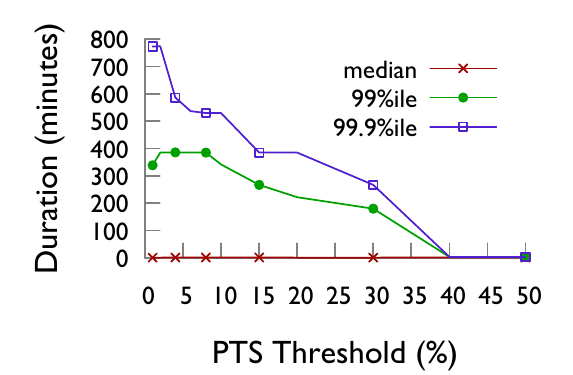}
	\caption{Gemini}
	\end{subfigure}
	\begin{subfigure}[c]{0.46\textwidth}
    \label{fig:aries_hotpsot}
	\includegraphics{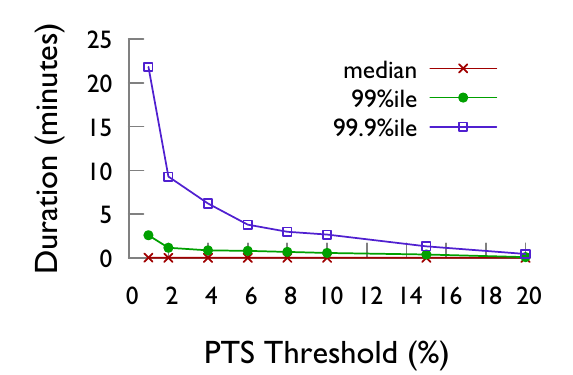}
	\caption{Aries}
	\end{subfigure}
	\caption{Congested link durations vs. PTS threshold for Blue Waters (Gemini) and Edison (Aries)}
	\label{fig:hotspot}
\end{figure*}
In this section, we describe the following two results obtained from Monet~\cite{MONET:NSDI:2020} tool on field-congestion data.
\begin{itemize}
    \item impact of routing algorithms on congestion (see Subsection~\ref{subsec:impact_routing})
    \item impact of heterogeneity in link-bandwidth on congestion (see Subsection~\ref{subsec:impact_bandwidth})
\end{itemize}

\subsection{Impact of Routing Algorithms}
\label{subsec:impact_routing}
Figure~\ref{fig:hotspot} shows the quantile values for different congested link durations, i.e., durations for which the \emph{PTS} value on the link is above a fixed threshold ($PTS_{th}$). The figure leads to the following insights:
\begin{itemize}
\item \textit{Use of the dragonfly topology and
adaptive routing has led to improvement in congestion control between two generations of Cray interconnects.} The Dragonfly topology used in Aries has a low global diameter of one hop, which helps to contain the back pressure of congested links. Furthermore, adaptive routing allows packets to take a longer but less congested path, which helps to alleviate congestion on the minimal path. Figure~\ref{fig:hotspot} provides empirical evidence for that observation. For every $PTS_{th}$ threshold, the congested link duration in Aries is an order of magnitude less than in Gemini. For example, if the threshold for congestion is fixed at 15\% \emph{PTS}, while the median duration is close to zero in both systems, the 99.9th percentile duration is approximately 1 minute for Edison and 400 minutes for Blue Waters. However, while Aries manages long bouts of congestion better than Gemini does, application runtime variability due to network performance remains a concern\cite{wang2014performance}.

\item \textit{Detection of long-duration congestion using traffic measurements can facilitate intervention such as rank remapping or rescheduling of bully jobs~\cite{yang2016watch}}. The 99.9th percentile congested link duration observed in both systems for $PTS_{th} \le 20\%$ is greater than a minute. Such long duration congestion allows us to tolerate greater latency for detection and diagnosis in real time. Moreover, a diagnosis can be converted to actionable feedback to be used by tools such as TopoMesh \cite{TopoLbIPDPS06}, which can remap MPI ranks or the scheduler to reschedule bully jobs.
\end{itemize}

\begin{figure*}[ht]
	\centering
	\hspace{0.03\textwidth}
	\begin{subfigure}[c]{0.31\textwidth}
    \label{fig:gemini_hotspot_X}
	\includegraphics{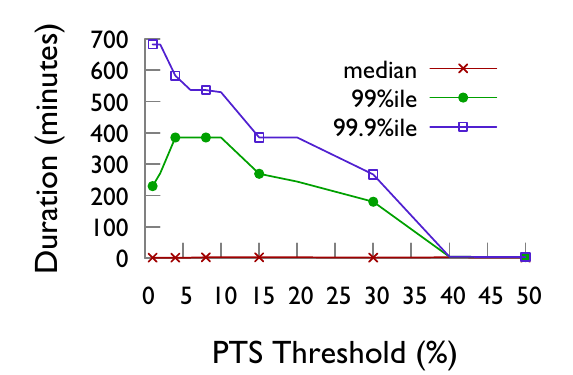}
	\caption{X+ and X-}
	\end{subfigure}
	\begin{subfigure}[c]{0.31\textwidth}
    \label{fig:gemini_hotspot_Y}
	\includegraphics{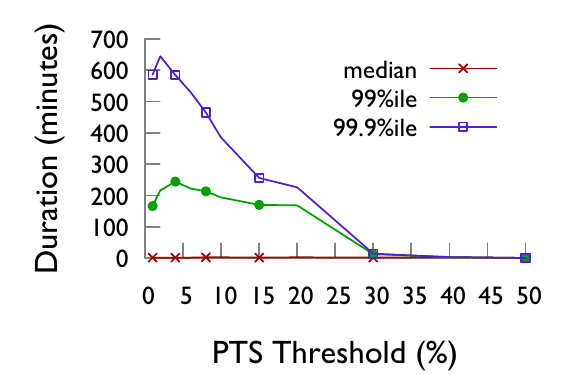}
	\caption{Y+ and Y-}
	\end{subfigure}
		\begin{subfigure}[c]{0.31\textwidth}
    \label{fig:gemini_hotspot_Z}
	\includegraphics{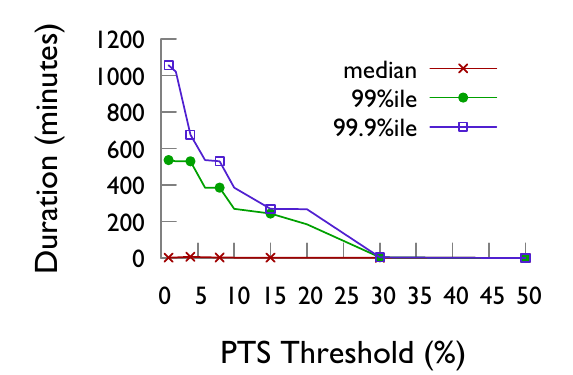}
	\caption{Z+ and Z-}
	\end{subfigure}
	\caption{Congested link durations for different link types in Gemini}
	\label{fig:gemini_hotspot_all}
\end{figure*}
\begin{figure*}[ht]
	\centering
	\hspace{0.03\textwidth}
	\begin{subfigure}[c]{0.31\textwidth}
    \label{fig:aries_hotspot_green}
	\includegraphics{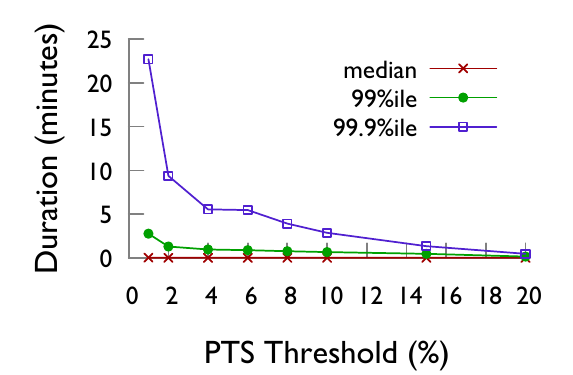}
	\caption{Green}
	\end{subfigure}
	\begin{subfigure}[c]{0.31\textwidth}
    \label{fig:aries_hotspot_black}
	\includegraphics{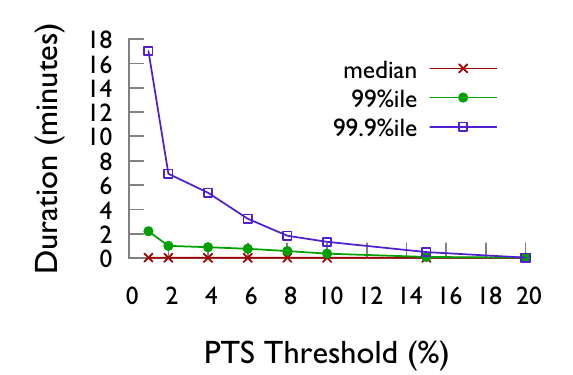}
	\caption{Black}
	\end{subfigure}
		\begin{subfigure}[c]{0.31\textwidth}
    \label{fig:aries_hotspot_blue}
	\includegraphics{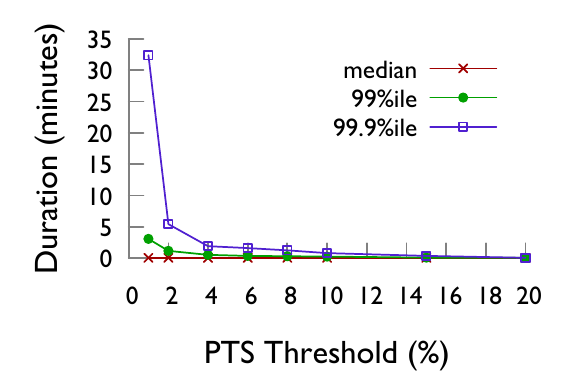}
	\caption{Blue}
	\end{subfigure}
	\caption{Congested link durations for different link types in Aries}
	\label{fig:aries_hotspot_all}
\end{figure*}
\subsection{Impact of Heterogeneity in Link-bandwidth}
\label{subsec:impact_bandwidth}
Heterogeneity in link bandwidth across different link types (electrical and optical links) increases the susceptibility to congestion.
Figure~\ref{fig:gemini_hotspot_all} (a), Figure~\ref{fig:gemini_hotspot_all} (b) and Figure~\ref{fig:gemini_hotspot_all} (c) respectively show congested link durations at different quantile values for \emph{X}, \emph{Y} and \emph{Z} directional links of Cray Gemini interconnect in Blue Waters, and Figure~\ref{fig:aries_hotspot_all} (a), Figure~\ref{fig:aries_hotspot_all} (b) and Figure~\ref{fig:aries_hotspot_all} (c) respecitvely show the congested link durations at different quantile values for \emph{Green}, \emph{Black} and \emph{Blue} links of Cray Aries interconnect in Edison. In Gemini, for higher $PTS_{th}$ thresholds ($\ge20\%$), links along the \emph{X} direction have longer lasting congestion than those on the \emph{Y} and \emph{Z} direction links. Similarly, in Aries, optical links (\emph{Blue}) have shorter and less severe bursts of congestion than the electrical links (\emph{Green} and \emph{Black}). \textit{Thus, mismatch and heterogeneity in link-bandwidth leads to varying levels of congestion along network path.}
\section{Conclusion}
\label{s:conclusion}
In this work, we demonstrated the use of Monet~\cite{MONET:NSDI:2020} to conduct long-term characterization of field-congestion data obtained from petascale systems across two different interconnect technologies: (i) Cray Gemini, which uses a 3-D torus topology, and (ii) Cray Aries, which uses the DragonFly topology. Future work will include an in-depth analysis of field-congestion data and methods to alleviate congestion issues in HPC interconnects.
\section{Acknowledgement*}
We thank  Larry Kaplan (Cray), Gregory Bauer (NCSA) and Jeremy Enos (NCSA) for having many insightful conversations. We thank K. Atchley and 
J. Applequist for their help in preparing the manuscript.

This material is based upon work supported by the U.S. Department of Energy, Office of Science, Office of Advanced Scientific Computing Research, under Award Number 2015-02674. This work is partially supported by NSF CNS 13-14891, and an IBM faculty award.

This research is part of the Blue Waters sustained-petascale computing project, which is supported by the National Science Foundation (awards OCI-0725070 and ACI-1238993) and the state of Illinois. Blue Waters is a joint effort of the University of Illinois at Urbana-Champaign and its National Center for Supercomputing Application. 

Sandia National Laboratories (SNL) is a multimission laboratory managed and operated by National Technology \& Engineering Solutions of Sandia, LLC, a wholly owned subsidiary of Honeywell International Inc., for the U.S. Department of Energy's National Nuclear Security Administration under contract DE-NA0003525. This paper describes objective technical results and analysis. Any subjective views or opinions that might be expressed in the paper do not necessarily represent the views of the U.S. Department of Energy or the United States Government.

This research used resources of the National Energy Research Scientific Computing Center (NERSC), a U.S. Department of Energy Office of Science User Facility operated under Contract No. DE-AC02-05CH11231.

{
    \bibliographystyle{IEEEtran}
    \bibliography{SNL,bibliography}
}

\end{document}